\documentstyle[12pt,a41]{article}
\newcommand{\beq}{\begin{equation}}
\newcommand{\eeq}{\end{equation}}
\newcommand{\bea}{\begin{eqnarray}}
\newcommand{\eea}{\end{eqnarray}}

\newcommand\ds{\displaystyle}

\newcommand{\Li}{\mbox{Li}}
\newcounter{lin}

\newcommand{\Sf}{\mbox{S}}

\begin{document}
\begin{titlepage}

\begin{flushleft}
DESY 97--160 \hfill {\tt hep-ph/9708388} \\
                    August 1997 \\
\end{flushleft}

\vspace{3cm}
\begin{center}
{\LARGE\bf On the Mellin Transform of the}

\vspace{3mm}
{\LARGE\bf Coefficient Functions of  \mbox{\boldmath $F_L(x,Q^2)$}}

\vspace{4cm}
{\large Johannes Bl\"umlein and Stefan Kurth}

\vspace{2cm}
{\large\it DESY--Zeuthen, Platanenallee 6, D--15735 Zeuthen, Germany}\\

\vspace{3cm}
\end{center}
\begin{abstract}
\noindent
The Mellin-transforms of the next-to-leading order Wilson coefficients
of the longitudinal structure function are evaluated.
\end{abstract}

\end{titlepage}

\section{Introduction}

\vspace{1mm}
\noindent
For the study of the scaling violations of deep-inelastic scattering
structure functions different techniques were developed~\cite{TEC,MEL}.
In most of the approaches the evolution equations of the parton
densities are solved as integro-differential equations in $x$-space
\cite{TEC}.
In the case of twist-2 operators the Mellin-transform  $M[f](N)$,
given by
\begin{equation}
\label{eq1}
M[f](N) =\int_0^1 dx x^{N-1} f(x)~,
\end{equation}
maps the convolutions  between the parton densities,
splitting functions, and coefficient functions into ordinary products,
which leads to a considerable simplification of the problem~\cite{MEL}.
Here $N$ denotes the integer moment-index. The analytic
continuation~\cite{TIT} of
the functional $M[f](N)$ into the complex plane is later used to perform
the inverse transformation to $x$-space.
It appears useful to apply the Mellin-representation not only for the
evolution equations of the parton densities but also to the
structure functions themselves. Their representation in $x$-space
is then obtained directly by a single inverse Mellin transform.
For various
next-to-leading order studies which were performed so far, as e.g.
of the structure functions $F_2(x,Q^2), g_1(x,Q^2)$, and $xF_3(x,Q^2)$,
the coefficient functions were requested only in $O(\alpha_s)$ and the
above procedure has been followed already.
In the case of the structure function $F_L(x,Q^2)$ the NLO
expressions~[4--6]  contain as well the Wilson
coefficients in $O(\alpha_s^2)$.
They were calculated in the $x$-representation in
ref.~\cite{cl2q} for the quarkonic and in ref.~\cite{cl2g1,cl2g2}
for the gluonic coefficient functions, where also the results for the
quarkonic coefficient functions of ref.~\cite{cl2q} were confirmed.
A careful check of all results
was performed in ref.~\cite{cl2g2} where a series of integer
moments was evaluated both numerically and analytically and compared
to the results of an independent calculation, ref.~\cite{LV}.
Previously the $x$-space expressions were used in various studies,
see~e.g.~\cite{ST}.

It is the aim of this note to shortly present the representation of the
coefficient functions in Mellin-$N$ space and their analytic continuation
to allow also to evaluate the longitudinal structure function at
next-to-leading order directly within the approach   used in
ref.~\cite{MEL}.
\section{The Wilson Coefficients in  $x$ space}
The longitudinal structure function $F_L(x,Q^2)$ has the representation
\begin{eqnarray}
\label{eqfl}
F_L(x,Q^2)\!=\!x\!\left \{
C_{\rm NS}(x,Q^2)\!\otimes\! f_{\rm NS}(x,Q^2)
           + \delta_f\!\left[C_{\rm S}(x,Q^2)\!\otimes\!\Sigma(x,Q^2)
           + C_{\rm g}(x,Q^2)\!\otimes\!G(x,Q^2)\right]\!\right\}
\end{eqnarray}
in the case of pure photon exchange.
The symbol $\otimes$ denotes the Mellin convolution
\begin{eqnarray}
\label{eqcon}
A(x,Q^2) \otimes B(x,Q^2) = \int_0^1 dx_1 \int_0^1 dx_2
\delta(x - x_1 x_2) A(x_1,Q^2) B(x_2,Q^2)~.
\end{eqnarray}
The combinations of parton densities are
\begin{eqnarray}
\label{eqpar}
f_{\rm NS}(x,Q^2) &=& \sum_{i=1}^{N_f} e_i^2 \left [q_i(x,Q^2)
+ \overline{q}_i(x,Q^2) \right], \\
\Sigma(x,Q^2) &=& \sum_{i=1}^{N_f}  \left [q_i(x,Q^2)
+ \overline{q}_i(x,Q^2) \right]~.
\end{eqnarray}
$G(x,Q^2)$ denotes the gluon density, $e_i$ the electric charge,
  and $\delta_f = (\sum_{i=1}^{N_f}
e_i^2)/N_f$, with  $N_f$  the number of active flavors.

The coefficient functions $C_i(x,Q^2)$ are given by
\begin{eqnarray}
\label{eqcoe}
C_{\rm NS}(z,Q^2) &=& a_s   c_{L,q}^{(1)}(z)+
                      a_s^2 c_{L,q}^{(2),NS}(z)  \nonumber\\
C_{\rm  S}(z,Q^2) &=& a_s^2 c_{L,q}^{(2), PS}(z) \\
C_{\rm  g}(z,Q^2) &=& a_s   c_{L,g}^{(1)}(z)+
                      a_s^2 c_{L, g}^{(2)}(z)~,  \nonumber
\end{eqnarray}
where $a_s = \alpha_s(Q^2)/(4\pi)$.
For convenience we list as well the coefficient functions in $x$-space,
since we will give the Mellin transforms of the individual contributing
functions separately below. The leading order coefficient functions
are given by~\cite{LO}
\begin{eqnarray}
\label{meq3}
 c_{L,q}^{(1)}(z) &=& 4 C_F z  \\
 c_{L,g}^{(1)}(z) &=& 8 N_f z (1-z)~.
\end{eqnarray}
In the $\overline{\rm MS}$ scheme the NLO coefficient functions read
[4--6]
\begin{eqnarray}
 c_{L,q}^{(2),NS}(z) &=& 4C_{F}(C_{A}-2C_{F})z
 \biggl\{4\frac{6-3z+47z^{2}-9z^{3}}{15z^{2}}\ln z \nonumber \\
& & -4\Li_{2}(-z)[\ln z -2\ln(1+z)]-8\zeta(3)
 -2\ln^{2}z[\ln(1+z)+\ln(1-z)] \nonumber \\
& & +4\ln z\ln^{2}(1+z)-4\ln z\Li_{2}(z) \
 +\frac{2}{5}(5-3z^{2})\ln^{2}z \nonumber \\
& &-4\frac{2+10z^{2}+5z^{3}-3z^{5}}
{5z^{3}}
[\Li_{2}(-z)+\ln z\ln(1+z)] \nonumber \\
& & +4\zeta(2)\left[\ln(1+z)+\ln(1-z)-\frac{5-3z^{2}}{5}\right]
 +8\Sf_{1,2}(-z)+4\Li_{3}(z) \nonumber \\
& &+4\Li_{3}(-z)-\frac{23}{3}\ln(1-z)
 -\frac{144+294z-1729z^{2}+216z^{3}}{90z^{2}}\biggr\} \nonumber \\
& & +8C_{F}^{2}z    \biggl\{\Li_{2}(z)+\ln^{2}z-2\ln z\ln(1-z)
 +\ln^{2}(1-z)-3\zeta(2)\nonumber \\
& & -\frac{3-22z}{3z}\ln z
 +\frac{6-25z}{6z}\ln(1-z)-\frac{78-355z}{36z}\biggr\}
\nonumber \\
& & -\frac{8}{3}C_{F}N_{f}z    \left\{2\ln z -\ln(1-z) -\frac{6-25z}
{6z}\right\},
\\
c_{L,q}^{(2),PS}(z) &=& \frac{16}{9z}C_{F}N_{f}\Bigl\{3(1-2z-2z^{2})
(1-z)\ln(1-z)
 +9z^{2}[\Li_{2}(z)+\ln^{2}z-\zeta(2)]\nonumber \\
& &~~~~ +9z(1-z-2z^{2})\ln z
 -9z^{2}(1-z)-(1-z)^{3}\Bigr\},
\\
\label{meq4}
c_{L,g}^{(2)}(z) &=& C_{F}N_{f}\biggl\{16z[\Li_{2}(1-z)
+\ln z\ln(1-z)] \nonumber \\
& & +\left(-\frac{32}{3}z+\frac{64}{5}z^{3}+\frac{32}{15z^{2}}\right)
[\Li_{2}(-z)+\ln z\ln(1+z)]
 +(8+24z-32z^{2})\ln(1-z) \nonumber \\
& & -\left(\frac{32}{3}z+\frac{32}{5}z^{3}\right)
\ln^{2}z +\frac{1}{15}\left(-104-624z+288z^{2}-\frac{32}{z}
\right)\ln z \nonumber \\
& & +\left(-\frac{32}{3}z+\frac{64}{5}z^{3}\right)\zeta(2)
-\frac{128}{15}-\frac{304}{5}z+\frac{336}{5}z^{2}+\frac{32}{15z}\biggr\}
\nonumber \\
& & +C_{A}N_{f}\biggl\{-64\Li_{2}(1-z)+(32z+32z^{2})[\Li_{2}(-z)
+\ln z\ln(1+z)] \nonumber\\
& &+(16z-16z^{2})\ln^{2}(1-z)
+(-96z+32z^{2})\ln z\ln(1-z)
\nonumber\\ & &
+\left(-16-144z+\frac{464}{3}z^{2}
+\frac{16}{3z}\right)\ln(1-z)
+48z\ln^{2}z    +(16+128z-208z^{2})\ln z
\nonumber\\  & &
+32z^{2}\zeta(2)+\frac{16}{3}
+\frac{272}{3}z-\frac{848}{9}z^{2}-\frac{16}{9z}\biggr\}~,
\end{eqnarray}
with $C_A = N_c =3, C_F = (N_c^2 - 1)/(2 N_c) = 4/3$.
The corresponding expressions in the DIS scheme are given
in~\cite{cl2g2}. The class of basic functions is the same for both
schemes.
\section{The Mellin Transform}

\vspace{1mm}
\noindent
For most of the functions $f_i(x)$ contributing to
eqs.~(\ref{meq3}--\ref{meq4}) the
Mellin transforms may be evaluated straightforwardly\footnote{Useful
relations can be found in refs.~\cite{YN,DD}.}. They
are listed in Table~1 for the individual functions.
In these cases infinite (or finite) sums occuring may be traced back
to representations  being based on  the functions
\begin{equation}
       \psi^{(k)}(z) =\frac{1}{\Gamma(z)}\frac{d^k}{d z^k} \Gamma(z)~,
\end{equation}
which represent the
sums
\begin{equation}
S_{n}(N) = \sum_{k=1}^{N}\frac{1}{k^{n}}~,
\end{equation}
with
\begin{equation}
S_{1}(N) = \psi(N+1)+\gamma_{E}~,
\end{equation}
and
\begin{equation}
S_{n}(N) = \frac{(-1)^{n-1}\psi^{(n-1)}(N+1)}{\Gamma(n)}+\zeta(n),
\; \; \; \; \; \;  n > 1~.
\end{equation}
Here  $\gamma_{E} = - \psi(1)$ is the Mascheroni constant and
$\zeta(n) = \sum_{k=1}^{\infty}(1/k^n)$ denotes the Riemann
$\zeta$-function.
One further may use Euler's relation
\begin{equation}
\Li_{2}(1-z) = -\Li_{2}(z)-\ln z\ln(1-z)+\zeta(2)
\end{equation}
to reduce the number of functions in the above expressions.
Similarly it is convenient to combine the functions
$\Li_{2}(-x)+\ln x\ln(1+x)$ whenever they occur with the same
coefficient   to
\begin{equation}
\label{meq2}
\Li_{2}(-x)+\ln x\ln(1+x) = -\frac{1}{2}\widetilde{\Phi}(x)
+\frac{1}{4}\ln^{2}x-\frac{\zeta(2)}{2}~,
\end{equation}
since the Mellin transform of the function
\begin{equation}
\widetilde{\Phi}(x) = \int_{x/(1+x)}^{1/(1+x)}\frac{dz}{z}\ln
\left(\frac{1-z}{z}\right)
\end{equation}
turns out to be simpler than that of either
$\Li_{2}(-x)$ or $\ln x\ln(1+x)$.
All these functions can be traced back to expressions which are built
solely out of the functions $\psi^{(k)}(z)$, see Table~1.

\pagebreak
\mbox{  }

\vspace*{-5mm}
\noindent
\begin{center}
\begin{tabular}{||r||c|l||}
\hline
\hline
& & \\ [-3mm]
\multicolumn{1}{||c||}{No.}&
\multicolumn{1}{c|}{${\ds f(z,r)}$}&
\multicolumn{1}{c||}{${\ds M[f](N)}$}\\
& & \\ [-3mm]
\hline\hline
& & \\ [-3mm]
1 & ${\ds z^{r}}$ & ${\ds \frac{1}{N+r}}$  \\
& & \\ [-3mm]
\hline
& & \\ [-3mm]
2 & ${\ds z^{r}\ln^{n}\!z}$ & ${\ds \frac{(-1)^{n}}{(N+r)^{n+1}}
\Gamma(n+1)}$  \\
& & \\ [-3mm]
\hline
& & \\ [-3mm]
3 & ${\ds z^{r}\ln(1-z)}$ & ${\ds -\frac{S_{1}(N+r)}{N+r}}$
\\
& & \\ [-3mm]
\hline
& & \\ [-3mm]
4    & ${\ds z^{r}\ln(1+z)}$ &
${\ds \frac{(-1)^{N+r-1}}{N+r}\biggl[-S_{1}(N+r)+\frac{1+(-1)^{N+r-1}}
{2}}$ ${\ds S_{1}\biggl(\frac{N+r-1}{2}\biggr)}$ \\
& & \\ [-3mm]
& & ${\ds +\frac{1-(-1)^{N+r-1}}{2}
S_{1}\biggl(\frac{N+r}{2}\biggr)\biggr]}$
${\ds +\Bigl[1+(-1)^{N+r+1}\Bigr]
\frac{\ln(2)}{N+r}}$  \\
& & \\ [-3mm]
\hline
& & \\ [-3mm]
5 & ${\ds z^{r}\ln^{2}\!(1-z)}$
 & ${\ds \frac{S_{1}^{2}(N+r)+S_{2}(N+r)}{N+r}}$
\\
& & \\ [-3mm]
\hline
& & \\ [-3mm]
6
 & ${\ds z^{r}\ln z\ln(1-z)}$ &
${\ds \frac{S_{1}(N+r)}{(N+r)^{2}}+\frac{1}{N+r}\Bigl[S_{2}(N+r)
-\zeta(2)\Bigr]}$  \\
& & \\ [-3mm]
\hline
& & \\ [-3mm]
7
 & ${\ds z^{r}\ln^{2}\!z\ln(1-z)}$ &
${\ds
  \frac{2}{N+r}\biggl[\zeta(3)+\frac{\zeta(2)}{N+r}
-\frac{S_{1}(N+r)}
{(N+r)^{2}}-}$
 ${\ds -\frac{S_{2}(N+r)}{N+r}-S_{3}(N+r)\biggr]}$  \\
& & \\ [-3mm]
\hline
& & \\ [-3mm]
8
 & ${\ds z^{r}\ln^{2}\!z\ln(1+z)}$
& ${\ds 2\frac{(-1)^{N+r}}{N+r}\biggl[\frac{S_{1}(N+r)}{(N+r)^{2}}
+\frac{S_{2}(N+r)}{N+r}+S_{3}(N+r)
 -\frac{\zeta(2)}{2(N+r)}-\frac{3\zeta(3)}{4}\biggr]}$  \\
& & \\ [-3mm]
& & ${\ds +\frac{1+(-1)^{N+r-1}}{2(N+r)}\biggl[\frac{2}{(N+r)^{2}}
S_{1}\left(\frac{N+r-1}{2}\right)
 +\frac{1}{N+r}S_{2}\left(
\frac{N+r-1}{2}\right)}$ \\
& & \\[-3mm]
& & ${\ds +\frac{1}{2}S_{3}\left(\frac{N+r-1}{2}\right)
 +\frac{4\ln 2}{(N+r)^{2}}\biggr]
 -\frac{1-(-1)^{N+r-1}}{2(N+r)}\biggl[\frac{2}{(N+r)^{2}}
S_{1}\left(\frac{N+r}{2}\right)}$ \\
& & \\[-3mm]
& & ${\ds +\frac{1}{N+r}S_{2}\left(\frac{N+r}{2}\right)
 +\frac{1}{2}S_{3}\left(\frac{N+r}{2}\right)\biggr]}$  \\
& & \\ [-3mm]
\hline
& & \\ [-3mm]
9
 & ${\ds z^{r}\Li_{2}(z)}$ & ${\ds \frac{1}{N+r}
  \biggl[\zeta(2)-\frac{S_{1}(N+r)}{N+r}\biggr]}$
  \\
& & \\ [-3mm]
\hline
& & \\ [-3mm]
10
 & ${\ds z^{r}\Li_{2}(z)\ln z}$ &
  ${\ds \frac{1}{(N+r)^{2}} \biggl[-2\zeta(2)+\frac{2S_{1}(N+r)}
  {N+r} +S_{2}(N+r)\biggr]}$  \\
& & \\ [-3mm]
& & \\
\hline \hline
\end{tabular}
\end{center}

\vspace*{3mm}
\noindent
{\sf Table~1~:  Mellin transforms of basic functions contributing to
the Wilson coefficients of $F_L(x,Q^2)$ in NLO.}
\renewcommand{\arraystretch}{1}
\begin{center}
\newpage
\noindent
\begin{tabular}{||r||c|l||}
\hline
\hline
& & \\ [-3mm]
\multicolumn{1}{||c||}{No.}&
\multicolumn{1}{c|}{${\ds f(z,r)}$}&
\multicolumn{1}{c||}{${\ds M[f](N)}$}\\
& & \\ [-3mm]
\hline\hline
& & \\ [-3mm]
11 & ${\ds z^{r}\widetilde{\Phi}(z)}$ &
${\ds
\frac{1}{(N+r)^{3}}+2\frac{(-1)^{N+r}}{N+r}
\biggl[S_{2}(N+r)-\zeta(2)\biggr]}$ \\
& & \\[-3mm]
& & ${\ds -\frac{1+(-1)^{N+r}}{2(N+r)}\biggl[S_{2}\left(
\frac{N+r}{2}\right)
-\zeta(2)\biggr]}$  \\
& & \\ [-3mm]
& & ${\ds +\frac{1-(-1)^{N+r}}{2(N+r)}\biggl[S_{2}\left(
\frac{N+r-1}{2}\right)
-\zeta(2)\biggr]}$  \\
& & \\ [-3mm]
\hline
& & \\ [-3mm]
12 & ${\ds z^{r}\Li_{2}(-z)\ln z}$ &
${\ds \frac{(-1)^{N+r-1}}{(N+r)^{2}}
\biggl[\frac{2S_{1}(N+r)}{N+r}+S_{2}(N+r)\biggr]}$  \\
& & \\ [-3mm]
& & ${\ds-\frac{1+(-1)^{N+r-1}}{2(N+r)^{2}}\biggl[\frac{2}{N+r}
S_{1}\left(\frac{N+r-1}{2}\right)
 +\frac{1}{2}S_{2}\left(
\frac{N+r-1}{2}\right)
+\frac{4\ln 2}{N+r}\biggr]}$  \\
& & \\ [-3mm]
& & ${\ds +\frac{1-(-1)^{N+r-1}}{2(N+r)^{2}}\biggl[\frac{2}{N+r}
S_{1}\left(\frac{N+r}{2}\right)
 +\frac{1}{2}S_{2}\left(\frac{N+r}{2}\right)
+\zeta(2)\biggr]}$
 \\
& & \\ [-3mm]
\hline
& & \\ [-3mm]
13
 & ${\ds z^{r}\Li_{3}(z)}$
& ${\ds \frac{1}{N+r}\biggl[\zeta(3)-\frac{\zeta(2)}{N+r}
  +\frac{S_{1}(N+r)}{(N+r)^{2}}\biggr]}$
\\
& & \\ [-3mm]
\hline
& & \\ [-3mm]
14 & ${\ds z^{r}\Li_{3}(-z)}$ &
${\ds (-1)^{N+r-1}\frac{S_{1}(N+r)}{(N+r)^{3}}
 -\frac{1+(-1)^{N+r-1}}{2(N+r)^{3}}\biggl[S_{1}\left(
\frac{N+r-1}{2}\right)+ 2\ln 2\biggr]}$  \\
& & \\ [-3mm]
& & ${\ds +\frac{1-(-1)^{N+r-1}}{2(N+r)^{3}}S_{1}\left(\frac{N+r}{2}
\right)+\frac{\zeta(2)}{2(N+r)^{2}}
 -\frac{3\zeta(3)}{4(N+r)}}$  \\
& & \\
\hline
\hline
\end{tabular}

\vspace*{3mm}
\noindent
{\sf Table~1~continued}
\renewcommand{\arraystretch}{1}
\end{center}

\vspace{5mm}
To the non-singlet coefficient function also the functions
\begin{equation}
\ln(x) \ln^2(1+x),~~{\rm S}_{1,2}(-x),~~{\rm and}~\Li_2(-x) \ln(1+x)
\nonumber
\end{equation}
contribute, which lead to different sums.
The Mellin transform of $\ln(x) \ln^2(1+x)$ can be obtained by a serial
expansion
\begin{eqnarray}
\label{ll12}
\int_{0}^{1}dx x^{N-1}\ln x\ln^{2}(1+x)
&=& \frac{2\gamma_{E}}{N^{2}}\biggl[\ln
  2-\frac{1}{2}\psi\left(\frac{N}{2}+1\right)+\frac{1}{2}
\psi\left(\frac{N+1}{2}\right)\biggr] \nonumber \\
&+&  \frac{\gamma_{E}}{2N}\biggl[\psi'\left(\frac{N}{2}+1\right)
-\psi'\left(\frac{N+1}{2}\right)\biggr]
-2\sum_{k=1}^{\infty}\frac{(-1)^{k}\psi(k)}{k(N+k)^{2}}~.
\end{eqnarray}
The infinite  sum in eq.~(\ref{ll12}) is easily evaluated by a recursive
algorithm. For numerical applications one may use as well
representations of $\log(1+x)$ in the range $x~\epsilon~[0,1]$,
cf.~\cite{HAST}, as, e.g.,
\begin{equation}
\label{meq1}
\ln(1+x) \simeq  \sum_{k=1}^{8}a_{k}x^{k}~,
\end{equation}
which are as accurate as $10^{-8}$. The coefficients $a_k$ read
\begin{eqnarray}
&&
a_1 = 0.9999964239~~~a_2 = -0.4998741238~~~a_3 = 0.3317990258~~~a_4 =
-0.2407338084 \nonumber\\
&&
a_5 = 0.1676540711~~~a_6 = -0.0953293897~~~a_7 = 0.0360884937~~~a_8 =
-0.0064535442~.  \nonumber
\end{eqnarray}
The Mellin transform of
$\ln x\ln^{2}(1+x)$ is then given by
\begin{equation}
\int_{0}^{1}dx x^{N-1}\ln x\ln^{2}(1+x) \simeq
-\sum_{k=2}^{16} \frac{b_k}{(N+k)^{2}}~,
\end{equation}
where the coefficients $b_{k}$ are obtained by taking the square
of the polynomial in eq.~(\ref{meq1}).

We furthermore observe that
${\rm S}_{1,2}(-x)$ and $\Li_2(-x) \ln(1+x)$
may be combined using the integral representation,
cf.~\cite{DD}\footnote{Note a misprint in eq.~(3.12.22) of ref.~\cite{DD}.
The factor in front of $\Li_3(-b/a)$ should be 2.},
\begin{equation}
\label{eq2}
F_1(x) = \frac{1}{x} \Bigl\{ \ln(1+x) \Li_2(-x) - \zeta(2)
+ {\rm S}_{1,2}(-x) - 2 \Li_3(-x) \Bigr \} =
\int_0^1 dy~\frac{\ln(y) \ln(1-y)}{1 + xy}~.
\end{equation}
Since the functions ${\rm S}_{1,2}(-x)$ and $\Li_{2}(-x)\ln(1+x)$ occur
with the same weight factor they can be delt with together with the help
of the
representation eq.~(\ref{eq2}).
In evaluating  the Mellin-transform  the $x$-integral
yields
\begin{equation}
\label{eq3}
\sum_{k=0}^{\infty} (-1)^k \frac{y^k}{N+k} \equiv \Phi(-y,1,N).
\end{equation}
Here the function $\Phi(y,a,b)$, ref.~\cite{PRUD}, is related to the
generalized Riemann $\zeta$-function $\zeta(c,n)$~\cite{PRUD1}.
Calculating the $y$-integral  by
using eq.~(6) of table~1, one finally
obtains
\begin{equation}
\label{eq42}
M[F_1](N) = \sum_{k=0}^{\infty} \frac{(-1)^k}{N+k} \left [
\frac{\psi(2+k) + \gamma_E}{(k+1)^2} - \frac{\psi'(2+k)}{k+1} \right].
\end{equation}
The representation eq.~(\ref{eq42}) is fastly converging since
for large values of $k$, $\psi(k) \sim \ln(k)$ and $\psi'(k) \sim 1/k$.
As in the case of the Mellin transforms of other functions emerging
in the $x$-space representation of the different coefficient and
splitting functions the poles in eqs.~(\ref{ll12}) and (\ref{eq42})
are situated at the non-positive integers.

We finally would like to comment on the analytic continuation of the
sums
\begin{equation}
\widetilde{S}_n(N) = \sum_{k=1}^{N}\frac{(-1)^{k}}{k^{n}} S_{1}(k)~,
~~~n \geq 2.
\end{equation}
which emerge in some of the Mellin-transforms.
$\widetilde{S}_2(N)$ contributes, e.g., to the NLO anomalous
dimensions. By using the relation
\begin{equation}
c_{k,n} =
\sum_{l=1}^{k}\frac{(-1)^l}{l^{n}} =
-\biggl\{\left(1-\frac{1}{2^{n-1}}\right)\zeta(n)
+\frac{(-1)^{k+n-1}}{2^{n}\Gamma(n)}\biggl[\psi^{(n-1)}
\left(\frac{k+1}{2}\right)-\psi^{(n-1)}\left(\frac{k}{2}+1\right)
\biggr]\biggr\}
\end{equation}
one may express $\widetilde{S}_k(N)$ as
\begin{eqnarray}
\label{eqSAn}
\widetilde{S}_n(N) &=&
(-1)^{N}\frac{S_{1}(N)}{N^{n}}
-\sum_{l=2}^{n}(-1)^{l}\zeta(l)
\biggl\{\left(1-\frac{1}{2^{n-l}}\right)\zeta(n-l+1) \nonumber \\
& &
-(-1)^N\frac{(-1)^{n-l}}{2^{n-l+1}\Gamma(n-l+1)}\biggl[\psi^{(n-l)}
\left(\frac{N  }{2}\right)-\psi^{(n-l)}\left(\frac{N+1}{2}\right)
\biggr]\biggr\} \nonumber \\
& & -       (-1)^n  \left[
\zeta(n)\ln 2 -(-1)^{N}\int_{0}^{1}dx
\frac{x^{N-1}\Li_{n}(x)}{1+x}
-\int_{0}^{1}dx\frac{\Li_{n}(x)}{1+x}\right],
\end{eqnarray}
where
the Mellin-transform of $\Li_n(x)/(1+x)$ is given by
\begin{eqnarray}
\label{eqLIN}
M[F_2](N) = \int_{0}^{1}dx\frac{\mbox{Li}_{n}(x)}{1+x}x^{N-1} &=&
\sum_{k=1}^{\infty}\frac{(-1)^{k}}{N+k} c_{k,n}~.
\end{eqnarray}
The series eq.~(\ref{eqLIN}) converges rather fast for $n>2$.
For the special case $n=2$, being delt with before in ref.~\cite{GRV},
one obtains from (\ref{eqSAn})
\begin{equation}
\widetilde{S}_2(N) =
-\frac{5}{8}\zeta(3)+(-1)^{N}\biggl[\frac{S_{1}(N)}{N^{2}}
-\frac{\zeta(2)}{2}G(N)+\int_{0}^{1}dx x^{N-1}\frac{\Li_{2}(x)}
{1+x}\biggr],
\end{equation}
with
\begin{equation}
\int_{0}^{1}dx\frac{\Li_{2}(x)}{1+x} = \zeta(2)\ln 2-\frac{5}{8}
\zeta(3)~~{\rm and}~~
G(N) = \psi\left(\frac{N+1}{2}  \right)-\psi\left(\frac{N}{2}\right)~.
\nonumber
\end{equation}
For $n=2$ the series eq.~(\ref{eqLIN}) converges  slowly since the
modulus of the  expansion coefficients $c_{k,2}$
approach $\zeta(2)/2 = \pi^2/12$ as
$k \rightarrow \infty$ and the series is
essentially  logarithmic. One may, however, rewrite $M[F_2](N)$ as
\begin{equation}
M[F_2](N) = \zeta(2) \ln(2) - \int_0^1 dx~x^{N-2}~\ln(1+x) \left [
(N-1) \Li_2(x) - \ln(1-x) \right].
\end{equation}
Again the function $\ln(1+x)$ can be represented by eq.~(\ref{meq1})
yielding
\begin{equation}
M[F_2](N) = \zeta(2) \ln(2) - \sum_{k=1}^8 a_k
\left\{ (N-1) \left[\frac{\zeta(2)}{N+k-1}   - \frac{
\psi(N+k)+\gamma_E}{(N+k-1)^2}\right] + \frac{\psi(N+k) + \gamma_E}
{N+k-1}  \right\}~.
\end{equation}
which holds at an accuracy of better than $4 \cdot 10^{-7}$
for $N~\epsilon~[1,20]$.
Another approximate expression for $M[F_2](N)$ was given in \cite{GRV}.

\vspace{2mm}
\noindent
{\bf Acknowledgement.}~We would like to thank W. van Neerven,
J. Vermaseren, F. Yndur\'ain, and A. Vogt for useful discussions.

\end{document}